\documentclass[aps,12pt,nofootinbib,showpacs]{revtex4}
\usepackage{graphics}
\usepackage{psfig}
\usepackage[dvips]{color}
\usepackage{amsmath}
\usepackage{amssymb}

\usepackage{epsfig}

\def\beq{\begin{equation}}
\def\eeq{\end{equation}}

\def\bea{\arraycolsep .1em \begin{eqnarray}}
\def\eea{\end{eqnarray}}
\def\Tr{{\rm Tr}}

\def\eq#1{(\ref{#1})}

\def\s0#1#2{\mbox{\small{$ \frac{#1}{#2} $}}}
\def\0#1#2{\frac{#1}{#2}}

\def\grgl{\:\hbox to -0.2pt{\lower2.5pt\hbox{$\sim$}\hss}{\raise3pt\hbox{$>$}}\:}
\def\klgl{\:\hbox to -0.2pt{\lower2.5pt\hbox{$\sim$}\hss}{\raise3pt\hbox{$<$}}\:}


\begin{document}

{\small\begin{flushright}
CERN-PH/TH-2005-257, SHEP-04-32, PITHA 05/20\\[10ex] \end{flushright}
}
\title{Fixed points of quantum gravity in extra dimensions\\[3ex]
}

\author{Peter Fischer${}^a$ and Daniel F.~Litim${}^{b,c}$}

\affiliation{\mbox{${}^a$  Institut f\"ur Theoretische Physik E,
RWTH Aachen,
D - 52056 Aachen}\\
 \mbox{${}^b$ School of Physics and Astronomy, U Southampton,
Highfield, SO17 1BJ, U.K.}\\
\mbox{${}^c$ Physics Department, CERN, Theory Division, CH -- 1211 Geneva 23.}
}%

\begin{abstract}
${}$\\[-1ex]

\centerline{\bf Abstract} We study quantum gravity in more than four
dimensions with renormalisation group methods. We find a non-trivial
ultraviolet fixed point in the Einstein-Hilbert action.  The fixed point
connects with the perturbative infrared domain through finite renormalisation
group trajectories. We show that our results for fixed points and related
scaling exponents are stable.  If this picture persists at higher order,
quantum gravity in the metric field is asymptotically safe.  We discuss
signatures of the gravitational fixed point in models with
low scale quantum gravity and compact extra dimensions.
\end{abstract}

\pacs{04.60.-m, 04.50.+h, 11.15.Tk, 11.25.Mj\\[-2ex]}

\pagestyle{plain}
\setcounter{page}{1}

\maketitle

The physics of gravitational interactions in more than four space-time
dimensions has received considerable interest in recent years.  The
possibility that the fundamental Planck mass -- within a higher dimensional
setting -- may be as low as the electroweak scale
\cite{Arkani-Hamed:1998rs,Antoniadis:1990ew,Randall:1999ee} has stimulated
extensive model building and numerous investigations aiming at signatures of
extra spatial dimensions ranging from particle collider experiments to
cosmological and astrophysical settings.  Central to these scenarios is that
gravity lives in higher dimensions, while standard model particles are often
confined to the four dimensional brane (although the latter is not crucial in
what follows).  In part, these models are motivated by string theory, where
additional spatial dimensions arise naturally \cite{Antoniadis:1998ig}. Then
string theory would, at least in principle, provide for a short distance
definition of these theories which presently have to be considered as
effective rather than fundamental ones.  In the absence of an explicit
ultraviolet completion, gravitational interactions at high energies including
low scale gravity can be studied with effective field theory or semi-classical
methods, as long as quantum gravitational effects are absent, or suppressed by
some ultraviolet cutoff of the order of the fundamental Planck mass, $e.g.$
\cite{Giudice:1998ck}.
\\[-1ex]

One may then wonder whether a quantum theory of gravity in the metric degrees
of freedom can exist in four and more dimensions as a cutoff-independent,
well-defined and non-trivial local theory down to arbitrarily small distances.
It is generally believed that the above requirements imply the existence of a
non-trivial ultraviolet fixed point under the renormalisation group, governing
the short-distance physics. The corresponding fixed point action then
provides a microscopic starting point to access low energy phenomena of
quantum gravity.  This ultraviolet completion 
does apply for quantum
gravity in the vicinity of two dimensions, where an ultraviolet fixed point
has been identified with $\epsilon$-expansion techniques
\cite{Weinberg,Gastmans:1977ad,Aida:1996zn}.  In the last couple of years, a
lot of efforts have been put forward to access the four-dimensional case, and
a number of independent studies have detected an ultraviolet fixed point using
functional and renormalisation group methods in the continuum
\cite{Reuter:1996cp,Souma:1999at,Lauscher:2001ya,Lauscher:2002mb,Reuter:2001ag,Percacci:2002ie,Litim:2003vp,Bonanno:2004sy,Litim2005,Forgacs:2002hz}
and Monte Carlo simulations on the lattice
\cite{Hamber:1999nu,Ambjorn:2004qm}.
\\[-1ex]

Continuity in the dimension suggests that a non-trivial fixed point -- if it
exists in four dimensions and below -- should persist at least in the vicinity
and above four dimensions.  Furthermore, the critical dimension of quantum
gravity -- the dimension where the gravitational coupling has vanishing
canonical mass dimension -- is two.  For any dimension above the critical one,
the mass dimension of the gravitational coupling is negative.  Hence, from a
renormalisation group point of view, four dimensions are not special. 
More generally, one expects that the local structure of quantum fluctuations,
and hence local renormalisation group properties of quantum theories of
gravity, are qualitatively similar for all dimensions above the critical one,
modulo topological effects in specific dimensions.
\\[-1ex]

In this Letter, we perform a fixed point search for quantum gravity in more
than four dimensions \cite{Litim:2003vp} (see also \cite{Reuter:2001ag}). An
ultraviolet fixed point, if it exists, should already be visible in the purely
gravitational sector, to which we confine ourselves. Matter degrees of freedom
and gauge interactions can equally be taken into account.  We employ a
functional renormalisation group based on a cutoff effective action $\Gamma_k$
for the metric field
\cite{Reuter:1996cp,Souma:1999at,Lauscher:2001ya,Lauscher:2002mb,Reuter:2001ag,Percacci:2002ie,Litim:2003vp,Bonanno:2004sy,Litim2005,Branchina:2003ek},
see \cite{ERG-Reviews} and \cite{Litim:1998nf} for
reviews in scalar and gauge theories.  In Wilson's approach, the functional
$\Gamma_k$ comprises momentum fluctuations down to the momentum scale $k$,
interpolating between $\Gamma_\Lambda$ at some reference scale $k=\Lambda$ and
the full quantum effective action at $k\to 0$. The variation of the effective
action with the cutoff scale ($t=\ln k$) is given by an exact functional flow
\begin{equation}\label{ERG}
\partial_t\Gamma_k=
\s012\Tr\frac{1}{\Gamma_k^{(2)}+R_k}\partial_t R_k\,.
\end{equation}
The trace is a sum over fields and a momentum integration, and $R_k$ is a
momentum cutoff for the propagating fields. The flow relates the change in
$\Gamma_k$ with a loop integral over the full cutoff propagator. By
construction, the flow \eq{ERG} is finite and, together with the boundary
condition $\Gamma_\Lambda$, defines the theory. In renormalisable theories,
the cutoff $\Lambda$ can be removed, $\Lambda\to\infty$, and
$\Gamma_\Lambda\to \Gamma_*$ remains well-defined for arbitrarily short
distances. In perturbatively renormalisable theories, $\Gamma_*$ is given by
the classical action, $e.g.$ in QCD.  In perturbatively non-renormalisable
theories, proving the existence (or non-existence) of a short distance limit
$\Gamma_*$ is more difficult.  In quantum gravity, the functional $\Gamma_*$
should at least contain those diffeomorphism invariant operators which display
relevant or marginal scaling in the vicinity of the fixed point.  A fixed
point action qualifies as a fundamental theory if it is connected with the
correct long-distance behaviour by finite
renormalisation group trajectories $\Gamma_k$.\\[-1ex]

The flow \eq{ERG} is solved by truncating $\Gamma_k$ to a finite set of
operators, which can systematically be extended. Highest reliability and best
convergence behaviour is achieved through an optimisation of the momentum
cutoff \cite{Litim:2000ci,Litim:2002cf,FischerLang,Pawlowski:2003hq}.  We
employ the Einstein-Hilbert truncation where the effective action, apart from
a classical gauge fixing and the ghost term, is given as
\begin{equation}\label{EHk}
\Gamma_k=
\0{1}{16\pi G_k}\int d^Dx \sqrt{g}\left[-R(g)+2\bar\lambda_k\right]\,.
\end{equation}
In \eq{EHk}, $g$ denotes the determinant of the metric field $g_{\mu\nu}$,
$R(g)$ the Ricci scalar, $G$ the gravitational coupling constant, and $\bar
\lambda$ the cosmological constant.  In the domain of classical scaling $G_k$
and $\bar\lambda_k$ are approximately constant, and \eq{EHk} reduces to the
conventional Einstein-Hilbert action in $D$ euclidean dimensions. The
dimensionless renormalised gravitational and cosmological constants are
\begin{equation}\label{glk}
\begin{array}{l}
g_k=k^{D-2}\, G_k\, \equiv k^{D-2}\, Z^{-1}_{N,k}\ \bar G
\\[1ex]
\lambda_k=\,k^{-2}\, \bar\lambda_k\ ,
\end{array}
\end{equation}
where $\bar G$ and $\bar\lambda$ denote the couplings at some reference scale,
and $Z_{N,k}$ the wave function renormalisation factor for the newtonian
coupling. Their flows are given by
\begin{equation}\label{dglk}
\begin{array}{l}   
\partial_t g\,       \equiv  \beta_g = \big[ D-2 + \eta_N  \big] g 
\\[1ex]
  \partial_t \lambda\,  \equiv   \beta_\lambda
\end{array}
\end{equation}
with $\eta_N(\lambda,g)=-\partial_t \ln Z_{N,k}$ the anomalous dimension of
the graviton. Fixed points correspond to the simultaneous vanishing of
\eq{dglk}. Explicit expressions for \eq{dglk} and $\eta_N$ follow from
\eq{ERG} by projecting onto the operators in \eq{EHk}, using background field
methods. We employ a momentum cutoff with the tensor structure of
\cite{Reuter:1996cp} (Feynman gauge) and optimised scalar cutoffs (see
below). For explicit analytical flow equations, see
\cite{Litim:2003vp,Litim2005}. The ghost wave function renormalisation is set
to $Z_{C,k}=1$.  Diffeomorphism invariance can be controlled by modified Ward
identities \cite{Reuter:1996cp}, similar to those employed for non-abelian
gauge theories \cite{Freire:2000bq}.  \\[-1ex]

\begin{figure}
\begin{center}
\unitlength0.001\hsize
\begin{picture}(500,750)
\put(10,470){\epsfig{file=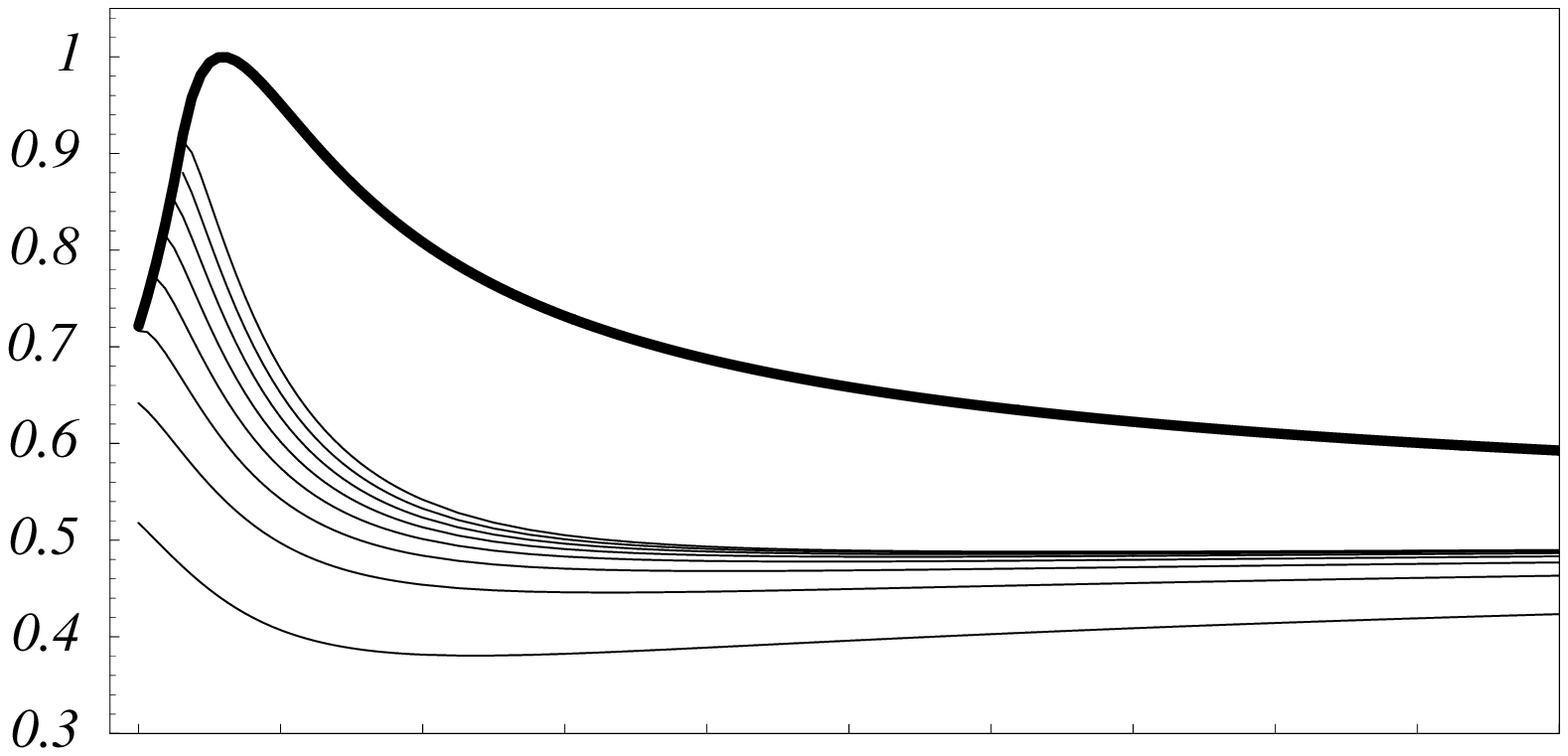,scale=0.469}}
\put(15,250){\epsfig{file=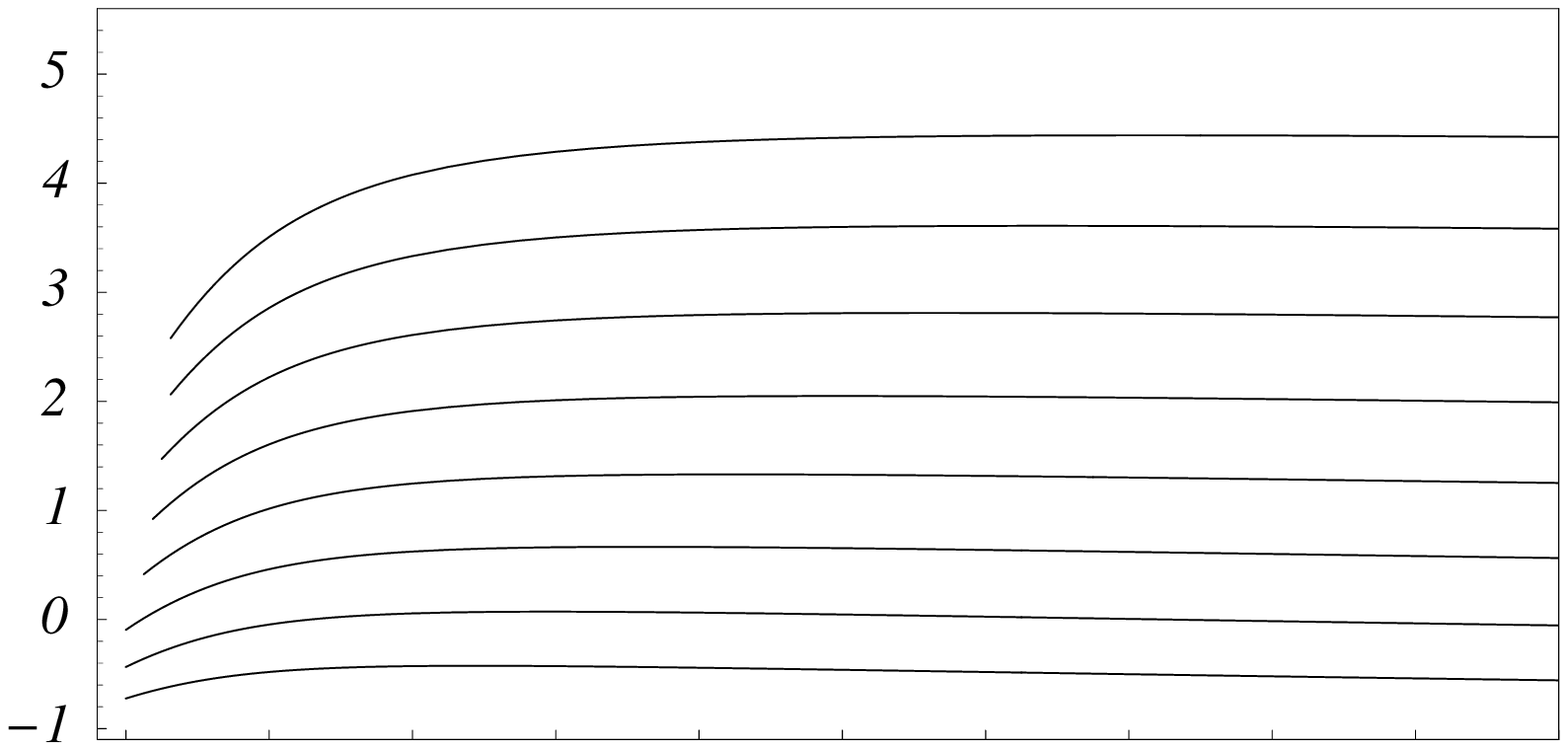,scale=0.463}}
\put(6,9){\epsfig{file=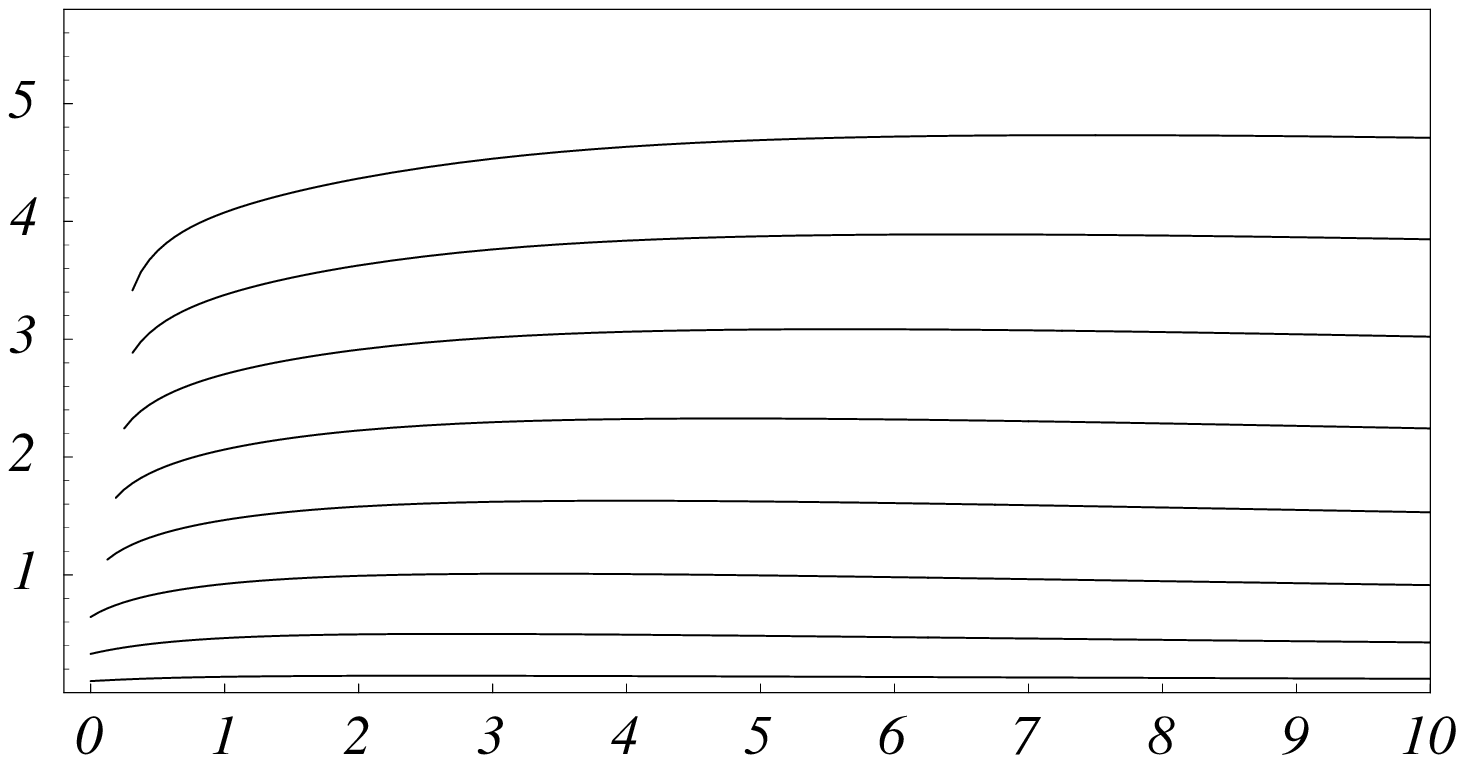,scale=0.502}}
\put(240,-20){$\log_{10}b$}
\put(120,660){a)$\ \lambda_*$}
\put(70,440){b)$\ \log_{10} g_*$}
\put(70,220){c)$\ \tau_*$}
\put(360,580){$\lambda_{\rm bound}$}
\end{picture}
\vspace{.5cm}
\caption{\label{pFP}
  Fixed points in $D=4+n$ dimensions with $n=0,...,7$
  (thin lines from bottom to top) as a function of the cutoff parameter $b$
  for momentum cutoff $r_{\rm mexp}$; a) the cosmological constant $\lambda_*$
  and $\lambda_{\rm bound}$ (thick line); b) the gravitational coupling $g_*$;
  c) the scaling variable $\tau_*=\lambda_*\, (g_*)^{2/(D-2)}$. }
\end{center}
\vspace{-.5cm}
\end{figure}

Two comments are in order. Firstly, the cosmological constant $\lambda$ obeys
$\lambda<\lambda_{\rm bound}$, where $2\lambda_{\rm bound}\equiv
\min_{q^2/k^2}[(q^2+R_k(q^2))/k^2]$ depends on the momentum cutoff $R_k(q^2)$
and $q^2\ge 0$ denotes (minus) covariant momentum squared.  Elsewise the flow
\eq{ERG}, \eq{dglk} could develop a pole at $\lambda=\lambda_{\rm bound}$.
The property $\lambda<\lambda_{\rm bound}$ is realised in any theory where
$\Gamma_k^{(2)}$ develops negative eigenmodes, and simply states that the
inverse cutoff propagator $\Gamma_k^{(2)} +R_k$ stays positive (semi-)
definite \cite{Litim:2006nn}.  Secondly, we detail the momentum cutoffs for
the numerical analysis.  We introduce $R_k(q^2)=q^2\, r(y)$, where
$y=q^2/k^2$.  Within a few constraints regulators can be chosen freely
\cite{Litim:2000ci}.  We employ $r_{\rm mexp}=b/((b+1)^{y}-1)$, $r_{\rm
  exp}=1/(\exp c y^b -1)$, $r_{\rm mod}=1/(\exp[c(y+(b-1)y^b)/b] -1)$, with
$c=\ln 2$, and $r_{\rm opt}=b(1/y-1)\theta(1-y)$.  These cutoffs include the
sharp cutoff $(b\to\infty)$ and asymptotically smooth Callan-Szymanzik type
cutoffs $R_k\sim k^2$ as limiting cases, and $b$ is chosen from $[b_{\rm
  bound},\infty]$, where $b_{\rm bound}$ stems from $\lambda<\lambda_{\rm
  bound}$ \cite{FischerLang}. The larger the parameter $b$, for each class,
the `sharper' the corresponding momentum cutoff.
\\[-1ex]

   \begin{table}
\begin{center}
\begin{displaymath}
\begin{array}{ccccccccc}
\ \ \tau_*&&&&&&&&
\\[.5ex]\hline\\[-2.5ex]
{}\ r_{\rm mexp}\ 
&\ 0.132\ &\ 0.461\ &\ 0.933\ &\ 1.502\ &\ 2.142\ 
&\ 2.834\ &\ 3.568\ &\ 4.336\ \\ 
{}\ r_{\rm exp}\ 
&  0.134  &  0.468 &0.946& 1.521  & 2.165& 2.858& 3.591& 4.356  \\ 
{}\ r_{\rm mod}\ 
& 0.135 & 0.469 & 0.946 & 1.521& 2.162& 2.853& 3.585& 4.348   \\ 
{}\ r_{\rm opt}\ 
&  0.137 &  0.478&  0.963&  1.544 &  2.192&  2.888&  3.623 &  4.389   \\[1ex] 
\hline
n&0&1&2&3&4&5&6&7\\[-2ex]
\end{array}
\end{displaymath}
\caption{Scaling variable $\tau_*$  in $D=4+n$ dimensions for
  various momentum cutoffs (see text).}
\label{tTau}
\end{center}
\vspace{-.7cm}
    \end{table}

\begin{figure}[t]
\begin{center}
  \unitlength0.001\hsize
\begin{picture}(500,900)
\put(0,690){\epsfig{file=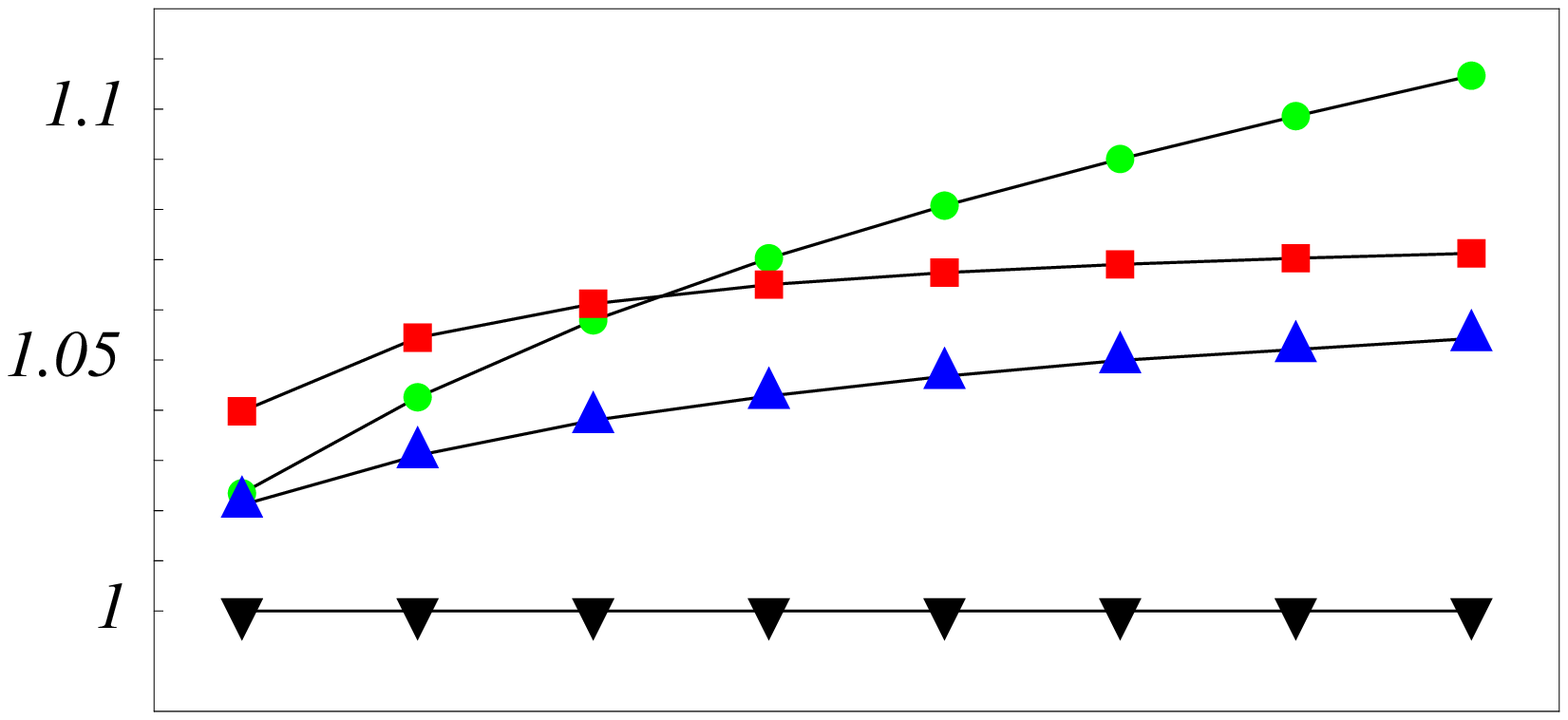,scale=0.45}}
\put(0,480){\epsfig{file=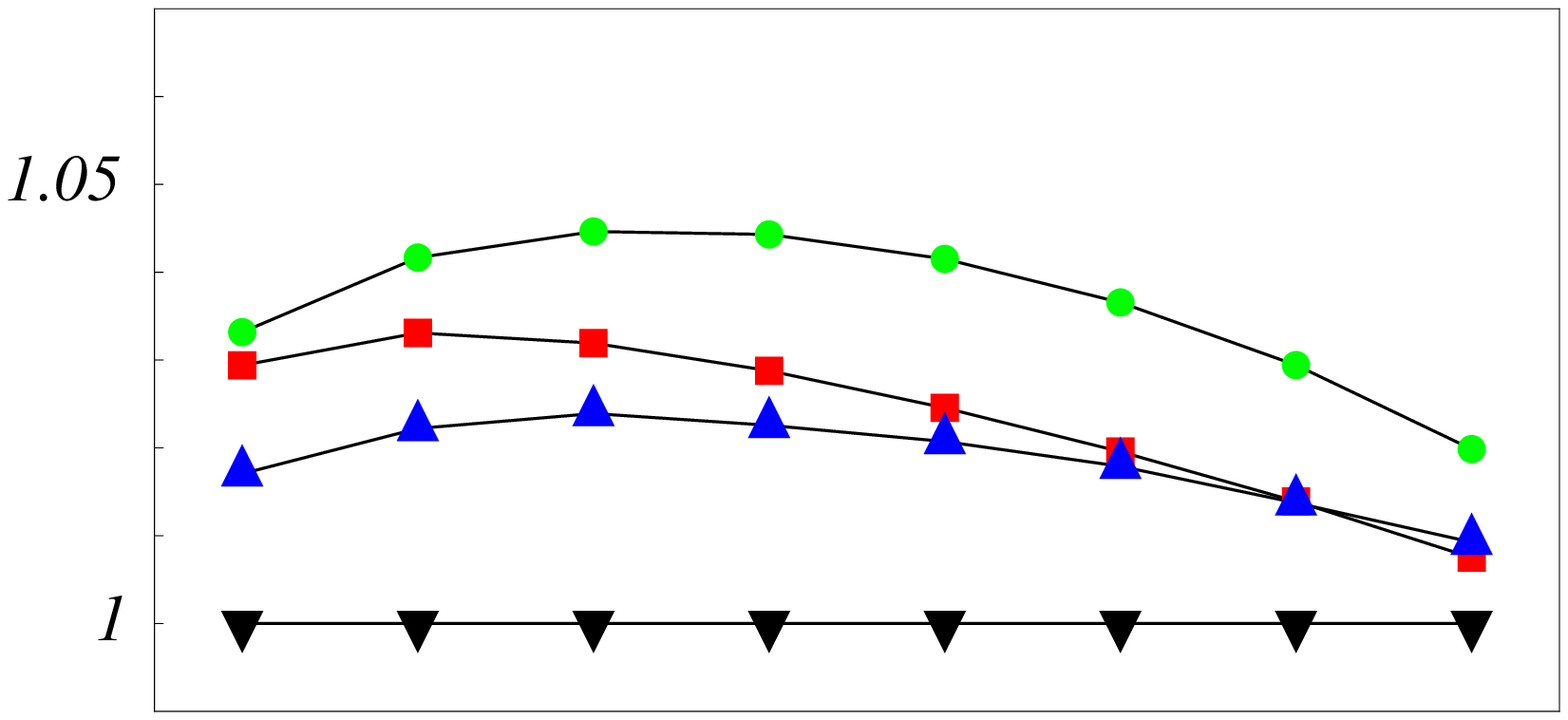,scale=0.451}}
\put(0,270){\epsfig{file=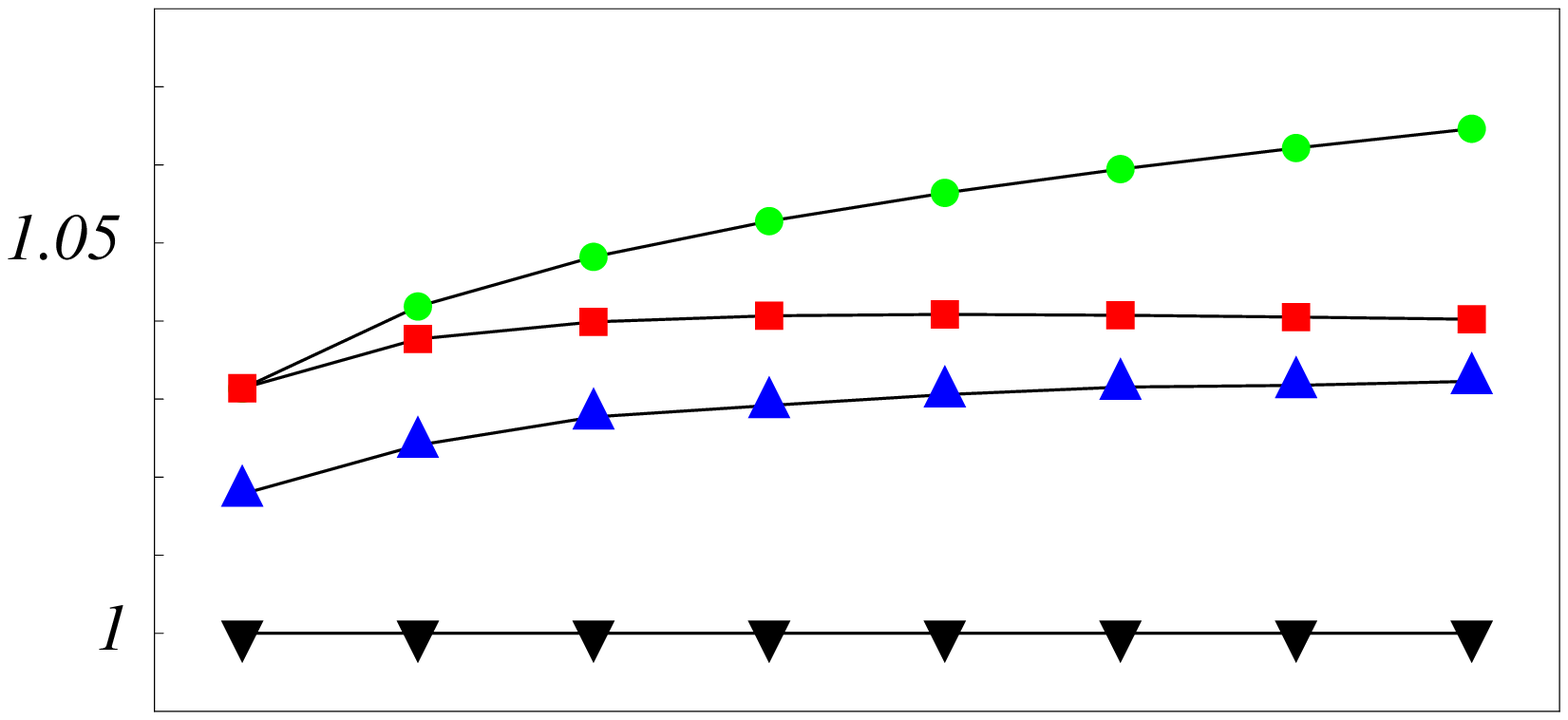,scale=0.45}}
\put(-33,30){\epsfig{file=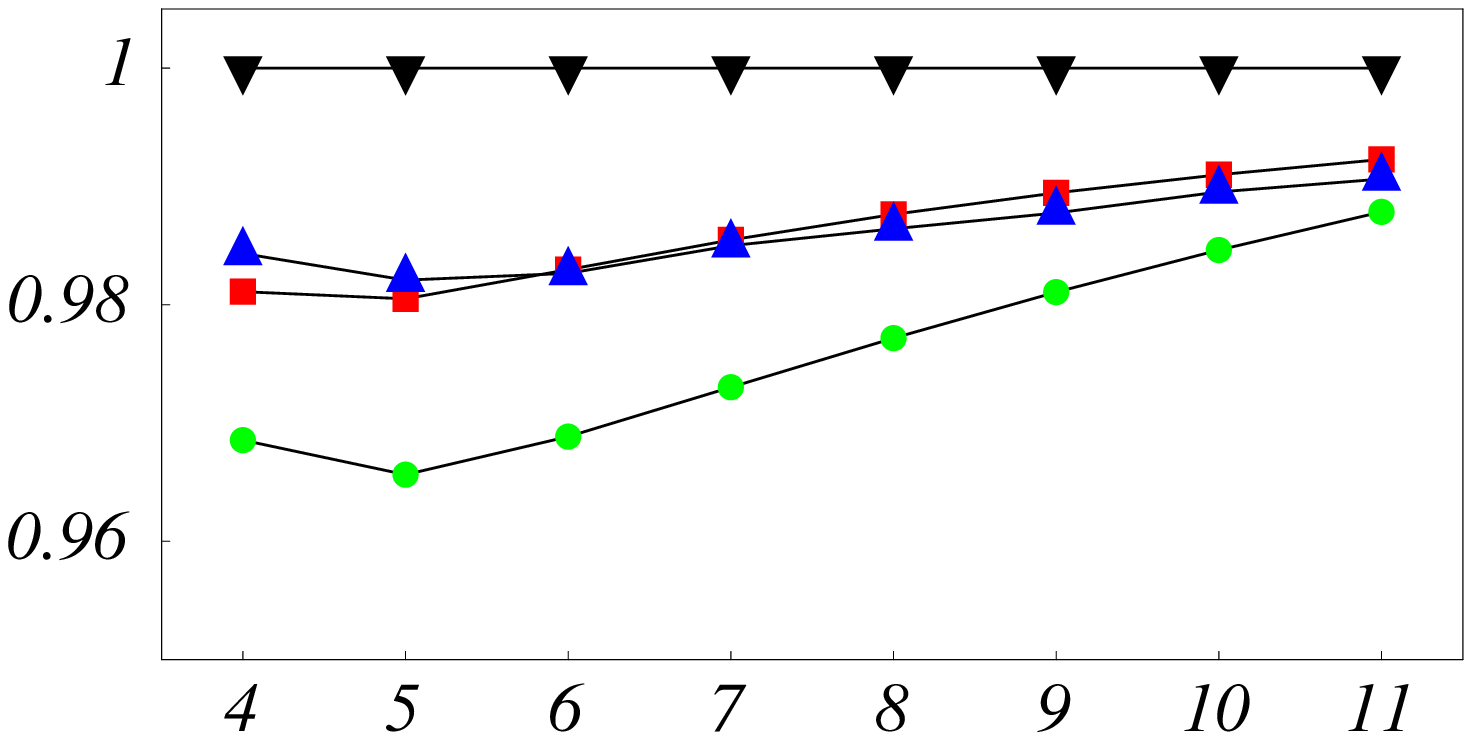,scale=0.512}}
\put(60,850){a)$\ \ \theta'/\theta'_{\rm opt}$}
\put(60,640){b)$\ \ \theta''/\theta''_{\rm opt}$}
\put(60,430){c)$\ \ |\theta|/|\theta|_{\rm opt}$}
\put(60,80){d)$\ \ \tau_*/\tau_{\rm opt}$}
\put(240,0){\large \bf $D$}
\end{picture}
\caption{\label{Vergleich}Comparison of $\theta'$, $\theta''$,
  $|\theta|$ and $\tau$ for different momentum cutoffs, normalised to the
  result for $r_{\rm opt}$ ($r_{\rm mexp}$ \textcolor{green}{$\bullet$},
  $r_{\rm exp}$ \textcolor{red}{$\blacksquare$}, $r_{\rm mod}$
  \textcolor{blue}{$\blacktriangle$}, $r_{\rm opt}$ $\blacktriangledown$).
  The relative variation, for all dimensions and all observables, is very
  small.}
\end{center}
\vspace{-.5cm}
\end{figure}

    Next we summarise our results for non-trivial ultraviolet fixed points
    $(g_*,\lambda_*)\neq(0,0)$ of \eq{dglk}, the related universal scaling
    exponents, trajectories connecting the fixed point with the perturbative
    infrared domain, the graviton anomalous dimension, cutoff independence,
    and the stability of the underlying expansion. We restrict ourselves to
    $D=4+n$ dimensions, with $n=0,...,7$ (see Figs.~\ref{pFP} and
    \ref{Vergleich}, and
    Tab.~\ref{tTau} -- \ref{tTheta}). \\[-1ex]
    
    {\it Existence.---} A real, non-trivial, ultraviolet fixed point exists
    for all dimensions considered, both for the cosmological constant and the
    gravitational coupling constant.  Fig.~\ref{pFP} shows our results for
    $\lambda_*$ and $\log_{10}g_*$ based on the momentum cutoff $r_{\rm mexp}$
    with parameter $b$ up to $10^{10}$. For small $b$, their numerical values
    depend strongly on $b$, while for large $b$, they become independent
    thereof.  Results similar to Fig.~\ref{pFP} are found for all momentum
    cutoffs indicated above \cite{FischerLang}. \\[-1ex]

    {\it Continuity.---} The fixed points $\lambda_*$ and $g_*$, as a function
    of the dimension, are continuously connected with their perturbatively
    known counterparts in two dimensions
    \cite{Souma:1999at,Lauscher:2001ya,Litim:2003vp,Litim2005}. \\ [-1ex]
    
    {\it Uniqueness.---} This fixed point is unique in all dimensions
    considered.
    \\ [-1ex]

{\it Positivity of the gravitational coupling.---} The gravitational coupling
constant only takes positive values at the fixed point.  Positivity is
required at least in the deep infrared, where gravity is attractive and
the renormalisation group running is dominated by classical scaling. Since the
flow $\beta_g$ in \eq{dglk} is proportional to $g$ itself, and the anomalous
dimension stays finite for small $g$, it follows that renormalisation group
trajectories cannot cross the line $g=0$ for any finite scale $k$. Therefore
the sign of $g$ is fixed along any trajectory, and positivity in the infrared
requires positivity already at an ultraviolet fixed point. At the critical
dimension $D=2$, the gravitational fixed point is degenerate with the gaussian
one $(g_*,\lambda_*)=(0,0)$, and, consequently, takes negative values below
two dimensions.
\\[-1ex]

    {\it Positivity of the cosmological constant.---} At vanishing $\lambda$,
    $\beta_\lambda$ is generically non-vanishing. Moreover, it depends on the
    running gravitational coupling. Along a trajectory, therefore, the
    cosmological constant can change sign by running through $\lambda=0$.
    Then the sign of $\lambda_*$ at an ultraviolet fixed point is not
    determined by its sign in the deep infrared. We find that the cosmological
    constant takes positive values at the fixed point, $\lambda_*>0$, for all
    dimensions and cutoffs considered. In pure gravity, the fixed point
    $\lambda_*$ takes negative values only in the vicinity of two dimensions.
    Once matter degrees of freedom are coupled to the theory, the sign of
    $\lambda_*$ can change, $e.g.$ in four dimensions \cite{Percacci:2002ie}.
    We expect this pattern to persist also in the higher-dimensional case.
    \\[-1ex]

   \begin{table}
\begin{center}
\begin{displaymath}
\begin{array}{ccccccccc}
\ \ \theta'&&&&&&&&
\\[.5ex]\hline\\[-2.5ex]
{}\ r_{\rm mexp}\ 
&\ 1.51\  &\ 2.80\ &\ 4.58\ &\ 6.71\ &\ 9.14\ &\ 11.9\ &\ 14.9\ &\ 18.2\ \\
{}\ r_{\rm exp}\ 
& 1.53& 2.83  & 4.60  & 6.68& 9.03  & 11.6&  14.5 & 17.6      \\
{}\ r_{\rm mod}\
& 1.51  & 2.77  & 4.50& 6.54& 8.86& 11.4&14.2&17.3      \\
{}\ r_{\rm opt}\  
& 1.48 & 2.69 & 4.33 & 6.27 & 8.46& 10.9& 13.5&16.4  \\[1ex]
\hline
n&0&1&2&3&4&5&6&7\\[-2ex]
\end{array}
\end{displaymath}
\caption{Scaling exponent $\theta'$ in $D=4+n$ dimensions for
  various momentum cutoffs (see text).}
\label{tTheta'}
\end{center}
\vspace{-.7cm}
    \end{table}
 
    {\it Dimensional analysis.---} In pure gravity (no cosmological constant
    term), only the sign of the gravitational coupling is well-defined, while
    its size can be rescaled to any value by a rescaling of the metric field
    $g_{\mu\nu}\to\ell\, g_{\mu\nu}$. In the presence of a cosmological
    constant, however, the relative strength of the Ricci invariant and the
    volume element can serve as a measure of the coupling strength. From
    dimensional analysis, we conclude that
\begin{equation}
\label{tau}
\tau_k =\bar\lambda_k \, (G_k)^{\frac{2}{D-2}}
\end{equation}
is dimensionless and invariant under rescalings of the metric field
\cite{Kawai:1992fz}.  Then the on-shell effective action is a function of
\eq{tau} only. In the fixed point regime, $\tau_k$ reduces to
$\tau_*=\lambda_*\, (g_*)^{2/(D-2)}$.  In Fig.~\ref{pFP}c), we have displayed
$\tau_*$ using the cutoff $r_{\rm mexp}$ for arbitrary $b$. In comparison with
the fixed point values in Figs.~\ref{pFP}a,b), $\tau_*$ only varies very
mildly as a function of the cutoff parameter $b$, and significantly less than
both $g_*$ and $\lambda_*$. This shows that \eq{tau} qualifies as a universal
variable in general dimensions. In Tab.~\ref{tTau}, we have collected our
results for \eq{tau} at the fixed point.  For all dimensions shown, $\tau_*$
displays only a very mild dependence on the cutoff function.
\\[-1ex]

{\it Universality.---} The fixed point values are non-universal. Universal
characteristics of the fixed point are given by the eigenvalues $-\theta$ of
the Jacobi matrix with elements $\partial_x \beta_y|_*$ and $x,y$ given by
$\lambda$ or $g$, evaluated at the fixed point. The Jacobi matrix is real
though not symmetric and admits real or complex conjugate eigenvalues.  For
all cases considered, we find complex eigenvalues $\theta\equiv\theta'\pm
i\theta''$.  In the Einstein-Hilbert truncation, the fixed point displays two
ultraviolet attractive directions, reflected by $\theta'>0$.  Complex scaling
exponents are due to competing interactions in the scaling of the volume
invariant $\int \sqrt{g}$ and the Ricci invariant $\int \sqrt{g}R$
\cite{Litim2005}. The eigenvalues are real in the vicinity of two dimensions,
and in the large-$D$ limit, where the fixed point scaling is dominated by the
$\int \sqrt{g}$ invariant \cite{Litim:2003vp}. $\theta'$ and $|\theta|$ are
increasing functions of the dimension, for all $D\ge 4$ \cite{Litim2005}.  For
the dimensions shown here, $\theta''$ equally increases with dimension.
\\[-1ex]
    
{\it UV-IR connection.---} A non-trivial ultraviolet fixed point is physically
feasible only if it is connected to the perturbative infrared domain by
well-defined, finite renormalisation group trajectories.  Elsewise, it would
be impossible to connect the known low energy physics of gravity with the
putative high energy fixed point. A necessary condition is $\lambda_* <
\lambda_{\rm bound}$, which is fulfilled.  Moreover, we have confirmed by
numerical integration of the flow that the fixed points are connected to the
perturbative infrared domain by well-defined trajectories.
\\[-1ex]

   \begin{table}
\begin{center}
\begin{displaymath}
\begin{array}{ccccccccc}
\ \ \theta'' &&&&&&&&
\\[.5ex]\hline\\[-2.5ex]
{}\ r_{\rm mexp}\ 
&\  3.14\  &\  5.37\  &\  7.46\  &\  9.46\  &\  11.4\ &\ 13.2\ &\ 14.9\ &\ 16.6\ \\  
{}\ r_{\rm exp}\ 
&  3.13  &  5.33  &  7.37  &  9.32  &  11.2  &  13.0  &  14.7  &  16.4 \\  
{}\ r_{\rm mod}\
&  3.10 &  5.27 &  7.31 &  9.26 &  11.1 &  13.0 &  14.7&  16.4   \\  
{}\ r_{\rm opt}\  
&  3.04 &  5.15&  7.14 &  9.05&  10.9 &12.7 &  14.5&  16.2  \\[1ex]  
\hline
n&0&1&2&3&4&5&6&7\\[-2ex]
\end{array}
\end{displaymath}
\caption{Scaling exponent $\theta''$ in $D=4+n$ dimensions for
  various momentum cutoffs (see text).}
\label{tTheta''}
\end{center}
\vspace{-.7cm}
    \end{table}

    {\it Anomalous dimension.---} The non-trivial fixed point implies a
    non-perturbatively large anomalous dimension for the gravitational field,
    due to \eq{dglk}, which takes negative integer values $\eta=2-D$ at the
    fixed point.\footnote{Integer anomalous dimensions are known from other
      gauge theories at a fixed point away from their canonical dimension,
      $e.g.$~abelian Higgs  \cite{Bergerhoff:1995zq} below or
      Yang-Mills \cite{Kazakov:2002jd} above four dimensions.}  The
    dressed graviton propagator ${\cal G}(p)$, neglecting the tensorial
    structure, is obtained from evaluating $1/(Z_{N,k}\,p^2)$ for momenta
    $k^2=p^2$. Then the graviton propagator scales as
\begin{equation}
\label{Gscaling}
{\cal G}(p)\sim 1/p^{2(1-\eta/2)}\,,
\end{equation} 
which reads $\sim 1/(p^2)^{D/2}$ in the deep ultraviolet and should be
contrasted with the $1/p^2$ behaviour in the perturbative regime. The
anomalous scaling in the deep ultraviolet implements a substantial suppression
of the graviton propagator.  We verified the crossover behaviour of the
anomalous dimension from perturbative scaling in the infrared to ultraviolet
scaling by numerical integration of the flow \eq{dglk}. More generally, higher
order vertex functions should equally display scaling characterised by
universal anomalous dimensions in the deep ultraviolet.  This is due to the
fact that a fixed point action $\Gamma_*$ is free of dimensionful parameters.
\\[-1ex]
  
    {\it Cutoff independence.---} Fixed points are found independently of the
    momentum cutoff, $e.g.$~Fig.~\ref{pFP}. The scaling exponents $\theta$,
    however, depend spuriously on $R_k$ due to the truncation.  This
    dependence strictly vanishes for the full, untruncated flow.  For best
    quantitative estimates of scaling exponents we resort to an optimisation,
    following \cite{Litim:2000ci,Litim:2002cf,FischerLang}, and use optimised
    values for $b$, for each class of cutoffs given above.  Optimised flows
    have best stability properties and lead to results closer to the physical
    theory \cite{Litim:2002cf}. In Tab.~\ref{tTau} -- \ref{tTheta}, we show
    our results for $\tau$, $\theta'$, $\theta''$ and $|\theta|$.  The
    variation in $\theta'$, $\theta''$, $|\theta|$ and $\tau$ is of the order
    of 11\%, 5\%, 7\% and 4\%, respectively (see Fig.~\ref{Vergleich}), and
    significantly smaller than the variation with $b$ \cite{FischerLang}. With
    increasing $n$, the variation slightly increases for $\theta'$ and
    $|\theta|$, and decreases for $\theta''$ and $\tau$.  The different
    dependences on the cutoff function, Fig.~\ref{Vergleich}a,c)
    vs.~Fig.~\ref{Vergleich}b,d), indicate that the observables are only
    weakly cross-correlated. The expected error due to the truncation \eq{EHk}
    is larger than the variation in Fig.~\ref{Vergleich}.  In this light, our
    results in the Einstein-Hilbert truncation are cutoff
    independent. \\[-1ex]

   \begin{table}
\begin{center}
\begin{displaymath}
\begin{array}{ccccccccc}
\ \ |\theta|  &&&&&&&&
\\[.5ex]\hline\\[-2.5ex]
{}\ r_{\rm mexp}\ 
&\ 3.49\ &\ 6.06\ &\ 8.76\ &\ 11.6\ &\ 14.6\ &\ 17.7\ &\ 21.1\ &\ 24.6\ \\
{}\ r_{\rm exp}\ 
& 3.49  & 6.03& 8.69& 11.5& 14.4 & 17.4& 20.7& 24.0 \\
{}\ r_{\rm mod}\
& 3.44  & 5.95& 8.58& 11.3 & 14.2& 17.3 & 20.5& 23.8 \\
{}\ r_{\rm opt}\  
&  3.38 &  5.81&  8.35&  11.0&  13.8&  16.8&  19.8&  23.1  \\[1ex]
\hline
n&0&1&2&3&4&5&6&7\\[-2ex]
\end{array}
\end{displaymath}
\caption{Scaling exponent $|\theta|$ in $D=4+n$ dimensions for
  various momentum cutoffs (see text).}
\label{tTheta}
\end{center}
\vspace{-.7cm}
    \end{table}

    {\it Convergence.---} The convergence of the results is assessed by
    comparing different orders in the expansion. The fixed point persists in
    the truncation where the cosmological constant is set to zero,
    $\lambda=\beta_\lambda=0$. Then, $\beta_g(g_*,\lambda=0)=0$ implies fixed
    points $g_*>0$ for all dimensions and cutoffs studied. The scaling
    exponent $\theta=-\partial\beta_g/\partial g|_*$ at $g_*$ is real and of
    the order of $|\theta|$ given in Tab.~\ref{tTheta}. The analysis can be
    extended beyond \eq{EHk}, $e.g.$~including $\int \sqrt{g}R^2$ invariants
    and similar.  In the four-dimensional case, $R^2$ interactions lead to a
    mild modification of the fixed point and the scaling exponents
    \cite{Lauscher:2001ya,Lauscher:2002mb}.  It is conceivable that the
    underlying
    expansion is well-behaved also in higher dimensions.  \\[-1ex]

In summary, we have found a unique and non-trivial ultraviolet fixed point
with all the right properties for quantum gravity in more than four
dimensions. Moreover, its universal characteristics are stable for the
truncations and cutoffs considered. Together with the renormalisation group
trajectories which connect the fixed point with the perturbative infrared
domain, our results provide a viable realisation of the asymptotic safety
scenario.  If the fixed point persists in extended truncations, quantum
gravity can well be formulated as a fundamental theory in the metric degrees
of freedom.
\\[-1ex]

As a first application, we discuss implications of the gravitational fixed
point for phenomenological models with compact extra dimensions, where gravity
propagates in a $D=4+n$ dimensional bulk. Under the assumption that standard
model particles do not spoil the fixed point, we can neglect their presence
for the following considerations. Without loss of generality, we consider $n$
extra spatial dimensions with compactification radius $L$.  The
four-dimensional Planck scale $M_{\rm Pl}$ is related to the $D$-dimensional
(fundamental) Planck mass $M_D$ and the radial length $L$ by the relation
$M^2_{\rm Pl}\sim M^2_{D} (M_D\,L)^{n}$, where $L^n$ is a measure for the
extra-dimensional volume.  A low fundamental Planck scale $M_D\ll M_{\rm Pl}$
therefore requires the scale separation
\begin{equation}
\label{ScaleSeparation}
1/L\ll M_D \,,
\end{equation} 
which states that the radius for the extra dimensions has to be much larger
than the fundamental Planck length $1/M_D$. For momentum scales $k\ll 1/L$,
where $\eta\approx 0$, the hierarchy \eq{ScaleSeparation} implies that the
running couplings scale according to their four-dimensional canonical
dimensions, with $G_k\approx$ const. At $k\approx 1/L$, the size of the extra
dimensions is resolved and, with increasing $k$, the couplings display a
dimensional crossover from four-dimensional to $D$-dimensional scaling. Still,
\eq{ScaleSeparation} implies that the graviton anomalous dimension stays small
and gravitational interactions remain perturbative.  This dimensional
crossover is insensitive to the fixed point in the deep ultraviolet.  In the
vicinity of $k\approx M_D$, however, the graviton anomalous dimension displays
a classical-to-quantum crossover from the gaussian fixed point $\eta\approx 0$
to non-perturbative scaling in the ultraviolet $\eta\approx 2-D$. This
crossover takes place in the full $D$-dimensional theory.  In the transition
regime, following \eq{Gscaling}, the propagation of gravitons is increasingly
suppressed, and the running gravitational coupling $G_k$ becomes very small
and approaches $g_*\, k^{2-D}$ with increasing $k$, as follows from \eq{glk}
and \eq{dglk}.  Therefore, the onset of the fixed point scaling cuts off
gravity-mediated processes with characteristic momenta at and above $M_D$, and
provides {\it dynamically} for an effective momentum cutoff of the order of
$M_D$. For momentum scales $\gg M_D$, gravity is fully dominated by the
non-perturbative fixed point and the associated scaling behaviour for vertex
functions.  Hence, the main new effects due to the fixed point set in at
scales about $M_D$. The significant weakening of $G_k$ and the dynamical
suppression of gravitons can be seen as signatures of the fixed point. This
behaviour affects the coupling of gravity to matter and could therefore be
detectable in experimental setups sensitive to the TeV energy range, $e.g.$~in
hadron colliders, provided that the fundamental scale of gravity is as low as
the electroweak scale. It will be interesting to identify physical observables
most sensitive to the above picture.
\\[-1ex]

{\it Acknowledgements:}\ We thank L.~Alvarez-Gaum\'e, I.~Antoniadis,
P.~Damgaard, C.~Jarlskog, W.~Kummer and G.~Veneziano for discussions.  PF
thanks the Austrian Ministerium f\"u{}r Bildung, Wissenschaft und Kultur for
support, and CERN for kind hospitality throughout main stages of this work.
DFL is supported by an EPSRC Advanced Fellowship.


\end{document}